\documentclass[aps,prl,twocolumn,showpacs,superscriptaddress]{revtex4-1}
\usepackage{graphicx}
\usepackage{color}

\begin{document}

\title{Non-magnetic and magnetic thiolate-protected Au25 superatoms on Cu(111), Ag(111) and Au(111)
surfaces}
\author{Xi Chen}
\affiliation{Department of Chemistry, Nanoscience Center, University of Jyv\"askyl\"a, 40014 Jyv\"askyl\"a, Finland}
\author{Mikkel Strange*}
\affiliation{Department of Chemistry, Nanoscience Center, University of Jyv\"askyl\"a, 40014 Jyv\"askyl\"a, Finland}
\author{Hannu H\"akkinen**}
\affiliation{Department of Chemistry, Nanoscience Center, University of Jyv\"askyl\"a, 40014 Jyv\"askyl\"a, Finland}
\affiliation{Department of Physics, Nanoscience Center, University of Jyv\"askyl\"a, 40014 Jyv\"askyl\"a, Finland}

\date{\today} 



\begin{abstract}
Geometry, electronic structure, and magnetic properties of methylthiolate-stabilized Au$_{25}$L$_{18}$ and MnAu$_{24}$L$_{18}$  (L = SCH$_3$) clusters adsorbed on noble-metal (111) surfaces have been investigated by using spin-polarized density functional theory computations. The interaction between the cluster and the surface is found to be mediated by charge transfer mainly from or into the ligand monolayer.  The electronic properties of the 13-atom metal core remain in all cases rather undisturbed as compared to the isolated clusters in gas phase. The Au$_{25}$L$_{18}$ cluster retains a clear HOMO - LUMO energy gap in the range of 0.7 eV to 1.0 eV depending on the surface. The ligand layer is able to decouple the electronic structure of the magnetic MnAu$_{24}$L$_{18}$ cluster from Au(111) surface, retaning a high local spin moment
of close to 5 $\mu_{B}$ arising from the spin-polarized Mn(3d) electrons. These computations imply that the thiolate monolayer-protected gold clusters may be used as promising building blocks with tunable energy gaps, tunneling barriers, and magnetic moments for applications in the area of
electron and/or spin transport. 

\end{abstract}


\maketitle

 Ligand monolayer protected gold clusters (MPCs) have received a lot of interest since mid-1990's due in part to
their enhanced stability, easy synthesis, rather well-defined monodispersity and potential applications in nanotechnology.
For example, the MPCs may be used as catalysts, sensors or building blocks for molecular electronic and spintronic devices
\cite{MurrayRev1,Astruc,HakRev,MurrayRev2,JinRev,OlgaNatChem}. 
They also form an interesting class of materials where the evolution of the clusters' electronic structure and
"metallicity" from molecular to colloidal (bulk-like) regimes can be explored, e.g., by measuring the optical or
electrochemical response  with varying cluster sizes.  

Recent 
refinements of the original synthesis\cite{Brust} of MPCs by several groups have enabled monodisperse samples of a few 
particularly stable ÒmagicÓ compounds in 1 - 3 nm range, and a few have by now been 
determined up to molecular precision; these include Au$_{20}$(SR)$_{16}$ \cite{Au20}, Au$_{25}$(SR)$_{18}^{-1/0}$
\cite{AU25-02,AU25-03,Au25neutral}, Au$_{38}$(SR)$_{24}$ \cite{ChakiAu38,JinAu38}, Au$_{40}$(SR)$_{24}$ \cite{Au40}, Au$_{68}$(SR)$_{34}$ \cite{Au68}, 
Au$_{102}$(SR)$_{44}$ \cite{Science1}, and compounds around 144 Au atoms and 60 thiolates \cite{ChakiAu38,JinAu144,Murray}.  
 The Au$_{25}$, Au$_{38}$ and Au$_{102}$ clusters are notable in this series since 
their total atomic structure has been determined from X-ray crystallography, opening the 
door to detailed theoretical analysis of the surface-covalent gold-sulphur bond and the 
electronic and geometric factors underlying the stability of these specific compounds.  

Up to now, these MPCs have been investigated as stand-alone entities computationally.
However, understanding the interaction of MPCs with their environments is
fundamentally important in order to assess their suitability as building blocks for components
in applications in molecular electronics or spintronics. Here, we have 
embarked on these studies by studying computationally the interaction of the
Au$_{25}$(SR)$_{18}$ cluster and its magnetically doped derivative MnAu$_{24}$(SR)$_{18}$ with a prototypical
metal surface, for which we select the noble metal (111).


\begin{figure}[tbp]
\includegraphics[width=1.0\linewidth,clip]{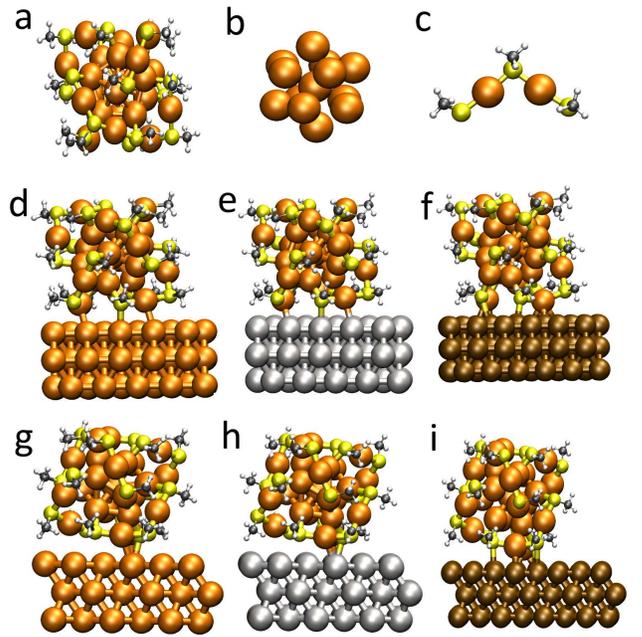}
\caption{ The structure of (a) Au$_{25}$L$_{18}^{-1}$ cluster, (b) Au$_{13}$ core, (c) Au$_2$L$_3$ unit and Au$_{25}$L$_{18}$ cluster on
(d) Au(111) , (e) Ag(111) and (f) Cu(111) surface. (g), (h), (i) are the same structures as (d), (e), (f), but rotated 90 degrees. The atoms are colored as: Au: orange, S: yellow, C: dark grey, H: light grey, Ag: silver, Cu: brown.}\label{fig:structures}
\end{figure}

A particularly stable phenylethanethiolate-protected gold cluster, Au$_{25}$(SCH$_2$CH$_2$Ph)$_{18}^{-1}$ was structurally characterized from X-ray crystallography in 2008 \cite{AU25-02,AU25-03} and its electronic and optical properties have been analyzed by using the density functional theory (DFT)  \cite{AU25-01, AU25-03, AU25-05}. The structure of the cluster is shown in Fig. \ref{fig:structures}a (from here on, we use methylthiolate SCH$_3$ as a model ligand and abbreviate L = SCH$_3$). It comprises an approximately icosahedral core of 13 Au atoms (see Fig. \ref{fig:structures}b) surrounded by six Au$_2$L$_3$ units (see Fig. \ref{fig:structures}c) in an octahedral arrangement with 12 S-Au   contacts that passivate the core surface. The cluster Au$_{25}$L$_{18}^{q}$ is known to be a stable redox species with at least three possible
charge (and spin) states:  $q$ = -1, 0, +1. This has led to suggestions that replacement of one or more core Au atoms by
a magnetic atom such as Mn should lead to a "magnetic superatom"; theoretically this has been shown to be
the case for an isolated MnAu$_{24}$L$_{18}$\cite{Khanna,AU25-05}. The robust geometry and a well-understood electronic and magnetic structure of these clusters motivated us to use Au$_{25}$L$_{18}$ and MnAu$_{24}$L$_{18}$ to investigate how metallic surfaces will affect the structure and properties of MPCs and whether the thiolate ligands can decouple the delocalized superatom orbitals and local spins from the electronic structure of the surface.

We use spin-polarized DFT as implemented in the GPAW code \cite{GPAW1,GPAW2}. The DFT calculations are performed with the projector-augmented-wave (PAW) method in a real space grid (0.2 \AA$\;$grid spacing) using the Perdew-Burke-Ernzerhof (PBE) generalized-gradient approximation \cite{PBE}. We regard Cu($3d^{10}4s^1$), Ag($4d^{10}5s^1$), Au($5d^{10}6s^1$),
Mn($3d^54s^2$), S($2s^22p^4$), C($2s^22p^2$) and H($1s^1$) electrons as the valence. The PAW setups for all the metals 
include scalar-relativistic corrections.  Periodic boundary conditions are applied in all the dimensions.  The Au(111), Ag(111) and Cu(111) substrates are modeled with three layers and 15 \AA$\;$vacuum is used to separate the Au$_{25}$ cluster and the image slab in the z direction. The two bottom metallic layers are fixed but all other atoms are relaxed during the geometry optimization until forces are below 0.05eV/\AA. A (5x6) surface cell is constructed for Au(111) and Ag(111) surfaces while a (6x8) surface cell is constructed for Cu(111). A Monkhorst-Pack 2x2x1 k-point sampling and the Bader method\cite{Bader2} are used to analyze the electronic structure and the charge state of the cluster.

The relaxed structures of Au$_{25}$L$_{18}$ on Au(111), Ag(111) and Cu(111) surfaces are shown in Fig. \ref{fig:structures}d-\ref{fig:structures}i. The geometrical structures of Au$_{25}$L$_{18}$ on Au(111) and Ag(111) surfaces are very similar. The shortest distance between the cluster and the surface atoms is 2.66 \AA$\;$for the Au(111) surface and 2.69 \AA$\;$for the Ag(111) surface which is the distance from one S atom to the Au (Ag) surface. Besides the S atom, there are two Au ligand atoms close to Au(111) and Ag(111) surfaces at 2.92 \AA, 2.93 \AA$\;$from the Au(111) surface and 2.93 \AA, 2.98 \AA$\;$from the Ag(111) surface. These three connections stabilize the cluster on the surface. Although some SCH$_3$ units are rotated by the interaction with the surfaces, the overall shape of the metal frame of the cluster remains spherical as in the case of  Au$_{25}$L$_{18}^{-1}$ in the gas phase. 

The structure of Au$_{25}$L$_{18}$ on Cu(111)  is significantly  different from Au(111) and Ag(111). The shortest distance between the cluster and the surface is also the distance between one S atom and Cu(111) surface (2.49 \AA), but the symmetry of the cluster is severely broken. Two ligand Au atoms are displaced towards the surface: one stays approximately on the hollow position of the Cu(111) surface and another stays approximately on the bridge position. The shorter distance and the distortion of the cluster show that Au$_{25}$L$_{18}$ and Cu(111) surface have a strong interaction. This is reflected by the calculated adsorption energy $\Delta$E of the cluster on the surface, for which 
we find values of -1.15 eV for Cu(111), -0.77 eV for Ag(111), and -0.53 eV for Au(111). It should be noted that our values are most likely
slight underestimations of the true adsorption energies since the PBE functional does not inlude van der Waals -type (vdW) interactions. Anyway we expect  that the above-mentioned systematic behavior is not changed by the neglected vdW effects.

\begin{table}[tbp]
\begin{center}
\begin{tabular}{|c|c|c|c|}
\hline\hline         
        & Cluster  &Core & Ligands\\
\hline 
Au$_{25}$L$_{18}^{-1}$  & -1.00  & 0.31 & -1.31\\
\hline
Au$_{25}$L$_{18}$/Au(111)  &0.28  & 0.41& -0.13\\
\hline
Au$_{25}$L$_{18}$/Ag(111) & -0.42 & 0.34& -0.76\\
\hline 
Au$_{25}$L$_{18}$/Cu(111) & -0.33 &0.41&-0.74\\
\hline\hline
\end{tabular}
\end{center}
\caption{The calculated charge decomposition (by the Bader method) of Au$_{25}$L$_{18}^{-1}$ and for the cluster on Au(111), Ag(111) and Cu(111). The core consists of the approximatively icosahedral Au$_{13}$ and the ligand
layer consists of the six Au$_2$L$_3$ units. }
\label{table1}
\end{table}

\begin{figure}[tbp]
\includegraphics[width=1.0\linewidth,clip]{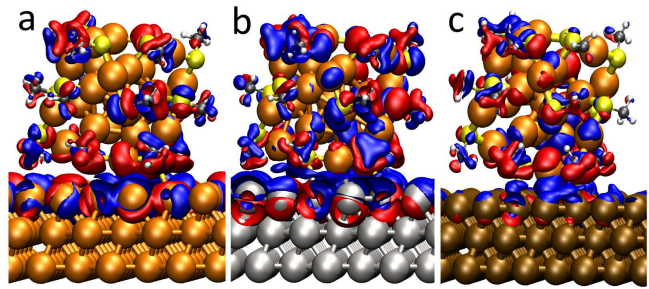}
\caption{ The isosurface (h=0.005\AA$^{-3}$) of $\delta \rho$ for Au$_{25}$L$_{18}$ on (a) Au(111), (b) Ag(111) and (c) Cu(111) surface. The blue and red regions show accumulated and depleted electron charge, respectively. The colors of atoms are as in Figure 1.}\label{fig:charge}
\end{figure}

When a MPC is placed on a metallic surface, the interactions between the MPC and the surface may induce the charge transfer effect which might change the electronic and magnetic properties of the cluster. In order to understand this effect, we calculated the Bader charge distribution of the Au$_{25}$L$_{18}$ cluster on the different surfaces and compared it to the stable Au$_{25}$L$_{18}^{-1}$ in gas phase.  The charge decomposition in the core and in the ligand layer of the cluster is  given in Table \ref{table1}. 0.28 electrons are transferred from  
Au$_{25}$L$_{18}$ to the Au(111) surface, while 0.42 electrons are transferred from the Ag(111) surface to the cluster and 0.33 electrons are transferred from the Cu(111) surface to the cluster. We note that the charge transfer between the cluster and the surface shows the same trend as the work function (W) of the surface. The order of the work functions is 
W(Ag(111))$<$W(Cu(111))$<$W(Au(111)).  If the Bader charges of the Au core and ligands are calculated separately, one can find out that the Au core has the similar charge state for Au$_{25}$L$_{18}^{-1}$ cluster and Au$_{25}$L$_{18}$ on the different surfaces, while the charge states of ligands are quite different, which indicates that the charge rearrangement of  
Au$_{25}$L$_{18}$ mainly happens in the ligand layer when the cluster absorbs on the Au(111), Ag(111) and Cu(111) surfaces.

The real-space charge redistribution can be visualized  by computing the quantity 
$\delta \rho=\rho_{clu+sub}-\rho_{clu}-\rho_{sub}$,
where $\rho_{clu+sub}$ is the electron charge density for the combined system while $\rho_{clu}$ and $\rho_{sub}$ are the electron charge densities of the noninteracting cluster and substrate, respectively.  The isosurface (h=0.005\AA$^{-3}$) of $\delta \rho$ for Au$_{25}$L$_{18}$ on Au(111), Ag(111) and Cu(111) is shown in Fig. \ref{fig:charge}. The blue region is where the electrons  are accumulated and the red one is where the electrons are depleted. It is clear that the charge rearrangement is localized mainly in the ligands of the cluster and in the first layer of the metal surface. The different adsorption mechanism can also be seen in this figure. For Au$_{25}$L$_{18}$/Au(111), the ligands which are close to the surface lose electrons, while the Au surface atoms which are close to ligands  gain electrons. The coulomb interaction between the positive charge of the cluster and the negative charge of the surface makes the cluster stable on the surface. For Au$_{25}$L$_{18}$/Cu(111), the electrons accumulate between the cluster and the surface which shows formation of a chemical-bond-like interaction between the cluster and the surface.

\begin{figure}[tbp]
\includegraphics[width=0.6\linewidth,clip]{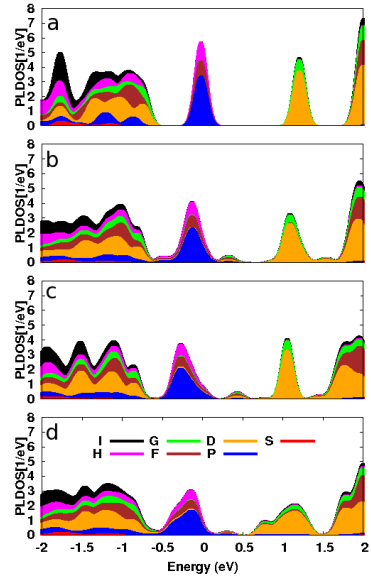}
\caption{ The angular-momentum-projected local density of states (PLDOS) 
(projection up to I symmetry, i.e $l=6$) for the Au$_{13}$ core of (a) Au$_{25}$L$_{18}^{-1}$, 
(b) Au$_{25}$L$_{18}$ on Au(111), (c) Au$_{25}$L$_{18}$ on Ag(111),
and (d) Au$_{25}$L$_{18}$ on Cu(111). The individual weights $c_{i,l}$ have been folded by a Gaussian
of 0.1 eV width. The Fermi energy is at zero.} \label{fig:YLM}
\end{figure}

It is of great interest to analyze the electronic structure of the adsorbed cluster,  especially the frontier orbitals near the Fermi level.
The Au$_{25}$L$_{18}^{-1}$ has been studied previously \cite{AU25-01, AU25-03, AU25-05}. The 
HOMO level is 3-fold degenerate whereas the LUMO is 2-fold degenerate with a HOMO-LUMO (HL) gap of about 1.3eV (PBE value). The character of the electronic states of the cluster can be analyzed by projection onto spherical harmonics placed at the mass center of the cluster \cite{PNAS1,AU25-01}:
\begin{equation}
c_{i,l}(R_0)=\sum_{m=-l}^l\int^{R_0}_0 r^2dr \left | \psi_{i,lm}(r)  \right|^2 ,
\end{equation} 
where \begin{equation}
\psi_{i,lm}(r) =\int d\hat{r} Y_{lm}(\hat{r}) \phi_i(r) ,
\end{equation}
Here $i$ is the index of the Kohn-Sham state $\phi(r)$ and Y$_{lm}$ is the spherical harmonic function with $l$ as the angular quantum number and $m$ as the magnetic quantum number. The angular momenta are considered up to $l=6$ (I-symmetry) and the expansion is made in a sphere of radius R$_{0}$, where R$_{0}$ is chosen to be 4.5 \AA. This analysis reveals the symmetry of the Kohn-Sham states of the system in the Au$_{13}$ core. Previously this 
analysis showed for Au$_{25}$L$_{18}^{-1}$ that the HOMO has P-symmetry and the LUMO has D-symmetry as shown in Fig. \ref{fig:YLM}a.
This result is easy to understand from the superatom model: The "free-electron" count for Au$_{25}$L$_{18}^{-1}$ 
is defined by the 26 Au(6s) electrons (including the extra electron from the negative charge) "donated" to the sp-subsystem, out of which 18 electrons are withdrawn by the thiolates, resulting in eight delocalized electrons with closed electronic shell at 1S$^2$1P$^6$.\cite{PNAS1,AU25-01}

Here we extend the use of eqs. 1 and 2 to analyze the electronic shells of the metal core of the adsorbed clusters.
The angular-momentum-projected local
electron density of states (PLDOS) for the Au$_{25}$L$_{18}$ on the Au(111), Ag(111) and Cu(111) are  shown in Fig. \ref{fig:YLM}b-\ref{fig:YLM}d respectively.
Neglecting the very small feature just above the Fermi level (originating  from the interaction between the cluster and the surface), the HOMO of the cluster in all cases has  a clear P-symmetry and the LUMO has a clear D-symmetry.
This shows that the "superatomic orbitals" exist also when the cluster is adsorbed on the metal surfaces. 
From the peak positions, the apparent "HL gaps" for the clusters can be determined.
For the Au(111) and Ag(111) surfaces, the HL gap is about 1.0 eV, slightly reduced from the
gas phase value of the Au$_{25}$L$_{18}^{-1}$ cluster.  The HL gap further diminishes when the cluster is on the Cu(111) surface (0.7eV), mainly because the LUMO states broaden into a wider band,  caused by the stronger interaction between the cluster and Cu surface. 

\begin{figure}[tbp]
\includegraphics[width=1.0\linewidth,clip]{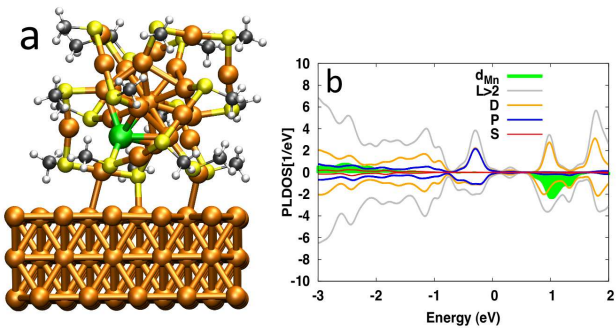}
\caption{ (a) The strucure of MnAu$_{24}$L$_{18}$ on Au(111). Mn is green, the other colors as in Figure 1. (b)
The angular-momentum-projected local density of states (PLDOS) 
of the MnAu$_{12}$ core  ($R_0=4.5$\AA\  in eq. 1) and Mn atom ($R_0=1.4 $ \AA\  in eq. 1)  of MnAu$_{24}$L$_{18}$ on Au(111). Top and bottom panels show the
PLDOS for spin up and down electrons, respectively. The individual weights $c_{i,l}$ have been folded by a Gaussian
of  0.1 eV width. The Fermi energy is at zero.  } \label{fig:Mn}
\end{figure}

If a cluster retains a  magnetic moment on the surface, it may have possible applications for 
spin-dependent transport in molecular spintronics. The neutral Au$_{25}$L$_{18}$  has a magnetic moment 1 $\mu_{B}$ in the gas phase, but we found that the moment is quenched on all  the (111) surfaces.  The  Mn-doped MnAu$_{24}$L$_{18}$  has a high moment of 5  $\mu_{B}$ in gas phase due to the localized Mn(3d$^5$) electrons.\cite{Khanna,AU25-05} We found that the energetically optimal site of the Mn atom is in the surface of the Au$_{13}$ core, in agreement with another recent DFT study\cite{Feng}. The energy differences to the next structure isomers are 0.30 eV and 0.36 eV (corresponding to Mn in the ligand layer and at the center of the core, respectively). Motivated by this result we replaced one of the Au$_{13}$ core surface atoms, facing Au(111), by the Mn atom. We find that the structure of the doped 
MnAu$_{24}$L$_{18}$/Au(111) and Au$_{25}$L$_{18}$/Au(111) are very similar (see fig. \ref{fig:Mn}a).
However in contrast to the Au$_{25}$L$_{18}$/Au(111) case, we observe that the magnetic moment of 
MnAu$_{24}$L$_{18}$/Au(111) remains very high, 4.9 $\mu_{B}$,  with 4.6 $\mu_{B}$ localized on the Mn atom. We show the PLDOS 
of the MnAu$_{12}$ core and the Mn atom of MnAu$_{24}$L$_{18}$/Au(111)  in Fig. \ref{fig:Mn}b. The 3d orbitals of the Mn atom remain fully spin-polarized which gives rise to the high magnetic moment of the cluster.  Simultaneously, the frontier orbitals of the cluster retain the P symmetry for HOMO and D symmetry for LUMO.

In conclusion, using methylthiolate protected Au$_{25}$L$_{18}$ and a doped MnAu$_{24}$L$_{18}$ as prototype clusters, we studied the interaction between MPCs and noble metal (111) surfaces. The geometric structure of the cluster and the adsorption mechanism  
depend on the interaction between the ligands and the surface, but the structure and the electronic properties  of the core remain.  The charge rearrangement is localized in the ligand layer which protects the overall electronic shell structure of both Au$_{25}$L$_{18}$ and MnAu$_{24}$L$_{18}$
and localized  spins of Mn 3d electrons in the core of MnAu$_{24}$L$_{18}$ by decoupling them from  the surface.
The HOMO-LUMO gap of the adsorbed cluster 
depends slightly on the metal, being the smallest on the Cu(111) which also has the strongest
interaction with the cluster. However a clear gap of about 0.7 eV to 1 eV is observed in all cases. 
Our results implicate that monodispersed thiolate-protected gold clusters retain their "superatom identity" also when adsorbed on prototypical metal surfaces. These novel building blocks can be experimentally synthesized in pure forms,
can be doped by magnetic atoms, and their ligand layer can be functionalized by varying the nature
(e.g., aryl- or alkylthiolates) and size of the ligands which affects the electronic coupling and tunneling barriers.
Utilizing this diversity  may open new ways to engineer building blocks for molecular electronic and spintronic devices.  

This research is supported by the Academy of Finland (FiDiPro project)  and CSC -  the Finnish IT Center for Science. X.C. wishes to thank O. Lopez-Acevedo and  P.A. Clayborne for help with GPAW.

 *Present address: Center for Atomic-scale Materials Design, Department of Physics, Technical University of Denmark, DK-2800 Kgs. Lyngby, Denmark.
 
 **Corresponding author: hakkinen.hannu@gmail.com

\bibliography{biblio}		

\begin{thebibliography}{27}%
\makeatletter
\providecommand \@ifxundefined [1]{%
 \@ifx{#1\undefined}
}%
\providecommand \@ifnum [1]{%
 \ifnum #1\expandafter \@firstoftwo
 \else \expandafter \@secondoftwo
 \fi
}%
\providecommand \@ifx [1]{%
 \ifx #1\expandafter \@firstoftwo
 \else \expandafter \@secondoftwo
 \fi
}%
\providecommand \natexlab [1]{#1}%
\providecommand \enquote  [1]{``#1''}%
\providecommand \bibnamefont  [1]{#1}%
\providecommand \bibfnamefont [1]{#1}%
\providecommand \citenamefont [1]{#1}%
\providecommand \href@noop [0]{\@secondoftwo}%
\providecommand \href [0]{\begingroup \@sanitize@url \@href}%
\providecommand \@href[1]{\@@startlink{#1}\@@href}%
\providecommand \@@href[1]{\endgroup#1\@@endlink}%
\providecommand \@sanitize@url [0]{\catcode `\\12\catcode `\$12\catcode
  `\&12\catcode `\#12\catcode `\^12\catcode `\_12\catcode `\%12\relax}%
\providecommand \@@startlink[1]{}%
\providecommand \@@endlink[0]{}%
\providecommand \url  [0]{\begingroup\@sanitize@url \@url }%
\providecommand \@url [1]{\endgroup\@href {#1}{\urlprefix }}%
\providecommand \urlprefix  [0]{URL }%
\providecommand \Eprint [0]{\href }%
\@ifxundefined \urlstyle {%
  \providecommand \doi  [0]{\begingroup \@sanitize@url \@doi}%
  \providecommand \@doi [1]{\endgroup \@@startlink {\doibase
  #1}doi:\discretionary {}{}{}#1\@@endlink }%
}{%
  \providecommand \doi  [0]{doi:\discretionary{}{}{}\begingroup
  \urlstyle{rm}\Url }%
}%
\providecommand \doibase [0]{http://dx.doi.org/}%
\providecommand \Doi [0]{\begingroup \@sanitize@url \@Doi }%
\providecommand \@Doi  [1]{\endgroup\@@startlink{\doibase#1}\@@Doi}%
\providecommand \@@Doi [1]{#1\@@endlink}%
\providecommand \selectlanguage [0]{\@gobble}%
\providecommand \bibinfo  [0]{\@secondoftwo}%
\providecommand \bibfield  [0]{\@secondoftwo}%
\providecommand \translation [1]{[#1]}%
\providecommand \BibitemOpen [0]{}%
\providecommand \bibitemStop [0]{}%
\providecommand \bibitemNoStop [0]{.\EOS\space}%
\providecommand \EOS [0]{\spacefactor3000\relax}%
\providecommand \BibitemShut  [1]{\csname bibitem#1\endcsname}%
\bibitem [{\citenamefont {Templeton}\ \emph {et~al.}(2000)\citenamefont
  {Templeton}, \citenamefont {Wuelfing},\ and\ \citenamefont
  {Murray}}]{MurrayRev1}%
  \BibitemOpen
  \bibfield  {author} {\bibinfo {author} {\bibfnamefont {A.~C.}\ \bibnamefont
  {Templeton}}, \bibinfo {author} {\bibfnamefont {W.~P.}\ \bibnamefont
  {Wuelfing}}, \ and\ \bibinfo {author} {\bibfnamefont {R.~W.}\ \bibnamefont
  {Murray}},\ }\href@noop {} {\bibfield  {journal} {\bibinfo  {journal} {Acc.
  Chem. Res.},\ }\textbf {\bibinfo {volume} {33}},\ \bibinfo {pages} {27}
  (\bibinfo {year} {2000})}\BibitemShut {NoStop}%
\bibitem [{\citenamefont {Daniel}\ and\ \citenamefont {Astruc}(2004)}]{Astruc}%
  \BibitemOpen
  \bibfield  {author} {\bibinfo {author} {\bibfnamefont {M.~C.}\ \bibnamefont
  {Daniel}}\ and\ \bibinfo {author} {\bibfnamefont {D.}~\bibnamefont
  {Astruc}},\ }\href@noop {} {\bibfield  {journal} {\bibinfo  {journal} {Chem.
  Rev.},\ }\textbf {\bibinfo {volume} {104}},\ \bibinfo {pages} {293} (\bibinfo
  {year} {2004})}\BibitemShut {NoStop}%
\bibitem [{\citenamefont {H{\"a}kkinen}(2008)}]{HakRev}%
  \BibitemOpen
  \bibfield  {author} {\bibinfo {author} {\bibfnamefont {H.}~\bibnamefont
  {H{\"a}kkinen}},\ }\href@noop {} {\bibfield  {journal} {\bibinfo  {journal}
  {Chem. Soc. Rev},\ }\textbf {\bibinfo {volume} {37}},\ \bibinfo {pages}
  {1847} (\bibinfo {year} {2008})}\BibitemShut {NoStop}%
\bibitem [{\citenamefont {Sardar}\ \emph {et~al.}(2009)\citenamefont {Sardar},
  \citenamefont {Funston}, \citenamefont {Mulvaney},\ and\ \citenamefont
  {Murray}}]{MurrayRev2}%
  \BibitemOpen
  \bibfield  {author} {\bibinfo {author} {\bibfnamefont {R.}~\bibnamefont
  {Sardar}}, \bibinfo {author} {\bibfnamefont {A.~M.}\ \bibnamefont {Funston}},
  \bibinfo {author} {\bibfnamefont {P.}~\bibnamefont {Mulvaney}}, \ and\
  \bibinfo {author} {\bibfnamefont {R.~W.}\ \bibnamefont {Murray}},\
  }\href@noop {} {\bibfield  {journal} {\bibinfo  {journal} {Langmuir},\
  }\textbf {\bibinfo {volume} {25}},\ \bibinfo {pages} {13840} (\bibinfo {year}
  {2009})}\BibitemShut {NoStop}%
\bibitem [{\citenamefont {Jin}(2010)}]{JinRev}%
  \BibitemOpen
  \bibfield  {author} {\bibinfo {author} {\bibfnamefont {R.~C.}\ \bibnamefont
  {Jin}},\ }\href@noop {} {\bibfield  {journal} {\bibinfo  {journal}
  {Nanoscale},\ }\textbf {\bibinfo {volume} {2}},\ \bibinfo {pages} {343}
  (\bibinfo {year} {2010})}\BibitemShut {NoStop}%
\bibitem [{\citenamefont {Lopez-Acevedo}\ \emph {et~al.}(2010)\citenamefont
  {Lopez-Acevedo}, \citenamefont {Kacprzak}, \citenamefont {Akola},\ and\
  \citenamefont {H{\"a}kkinen}}]{OlgaNatChem}%
  \BibitemOpen
  \bibfield  {author} {\bibinfo {author} {\bibfnamefont {O.}~\bibnamefont
  {Lopez-Acevedo}}, \bibinfo {author} {\bibfnamefont {K.~A.}\ \bibnamefont
  {Kacprzak}}, \bibinfo {author} {\bibfnamefont {J.}~\bibnamefont {Akola}}, \
  and\ \bibinfo {author} {\bibfnamefont {H.}~\bibnamefont {H{\"a}kkinen}},\
  }\href@noop {} {\bibfield  {journal} {\bibinfo  {journal} {Nature
  Chemistry},\ }\textbf {\bibinfo {volume} {2}},\ \bibinfo {pages} {329}
  (\bibinfo {year} {2010})}\BibitemShut {NoStop}%
\bibitem [{\citenamefont {Brust}\ \emph {et~al.}(1994)\citenamefont {Brust},
  \citenamefont {Walker}, \citenamefont {Bethell}, \citenamefont {Schiffrin},\
  and\ \citenamefont {Whyman}}]{Brust}%
  \BibitemOpen
  \bibfield  {author} {\bibinfo {author} {\bibfnamefont {M.}~\bibnamefont
  {Brust}}, \bibinfo {author} {\bibfnamefont {M.}~\bibnamefont {Walker}},
  \bibinfo {author} {\bibfnamefont {D.}~\bibnamefont {Bethell}}, \bibinfo
  {author} {\bibfnamefont {D.~J.}\ \bibnamefont {Schiffrin}}, \ and\ \bibinfo
  {author} {\bibfnamefont {R.}~\bibnamefont {Whyman}},\ }\href@noop {}
  {\bibfield  {journal} {\bibinfo  {journal} {Chem. Comm.},\ }\textbf {\bibinfo
  {volume} {1994}},\ \bibinfo {pages} {801} (\bibinfo {year}
  {1994})}\BibitemShut {NoStop}%
\bibitem [{\citenamefont {Zhu}\ \emph {et~al.}(2009)\citenamefont {Zhu},
  \citenamefont {Qian},\ and\ \citenamefont {Jin}}]{Au20}%
  \BibitemOpen
  \bibfield  {author} {\bibinfo {author} {\bibfnamefont {M.~Z.}\ \bibnamefont
  {Zhu}}, \bibinfo {author} {\bibfnamefont {H.~F.}\ \bibnamefont {Qian}}, \
  and\ \bibinfo {author} {\bibfnamefont {R.~C.}\ \bibnamefont {Jin}},\
  }\href@noop {} {\bibfield  {journal} {\bibinfo  {journal} {J. Am. Chem.
  Soc.},\ }\textbf {\bibinfo {volume} {131}},\ \bibinfo {pages} {7220}
  (\bibinfo {year} {2009})}\BibitemShut {NoStop}%
\bibitem [{\citenamefont {Heaven}\ \emph {et~al.}(2008)\citenamefont {Heaven},
  \citenamefont {Dass}, \citenamefont {White}, \citenamefont {Holt},\ and\
  \citenamefont {Murray}}]{AU25-02}%
  \BibitemOpen
  \bibfield  {author} {\bibinfo {author} {\bibfnamefont {M.~W.}\ \bibnamefont
  {Heaven}}, \bibinfo {author} {\bibfnamefont {A.}~\bibnamefont {Dass}},
  \bibinfo {author} {\bibfnamefont {P.~S.}\ \bibnamefont {White}}, \bibinfo
  {author} {\bibfnamefont {K.~M.}\ \bibnamefont {Holt}}, \ and\ \bibinfo
  {author} {\bibfnamefont {R.~W.}\ \bibnamefont {Murray}},\ }\href@noop {}
  {\bibfield  {journal} {\bibinfo  {journal} {J. Am. Chem. Soc},\ }\textbf
  {\bibinfo {volume} {130}},\ \bibinfo {pages} {3754} (\bibinfo {year}
  {2008})}\BibitemShut {NoStop}%
\bibitem [{\citenamefont {Zhu}\ \emph {et~al.}(2008){\natexlab{a}}\citenamefont
  {Zhu}, \citenamefont {Aikens}, \citenamefont {Hollander}, \citenamefont
  {Schatz},\ and\ \citenamefont {Jin}}]{AU25-03}%
  \BibitemOpen
  \bibfield  {author} {\bibinfo {author} {\bibfnamefont {M.}~\bibnamefont
  {Zhu}}, \bibinfo {author} {\bibfnamefont {C.~M.}\ \bibnamefont {Aikens}},
  \bibinfo {author} {\bibfnamefont {F.~J.}\ \bibnamefont {Hollander}}, \bibinfo
  {author} {\bibfnamefont {G.~C.}\ \bibnamefont {Schatz}}, \ and\ \bibinfo
  {author} {\bibfnamefont {R.}~\bibnamefont {Jin}},\ }\href@noop {} {\bibfield
  {journal} {\bibinfo  {journal} {J. Am. Chem. Soc},\ }\textbf {\bibinfo
  {volume} {130}},\ \bibinfo {pages} {5883} (\bibinfo {year}
  {2008}{\natexlab{a}})}\BibitemShut {NoStop}%
\bibitem [{\citenamefont {Zhu}\ \emph {et~al.}(2008){\natexlab{b}}\citenamefont
  {Zhu}, \citenamefont {Eckenhoff}, \citenamefont {Pintauer},\ and\
  \citenamefont {Jin}}]{Au25neutral}%
  \BibitemOpen
  \bibfield  {author} {\bibinfo {author} {\bibfnamefont {M.}~\bibnamefont
  {Zhu}}, \bibinfo {author} {\bibfnamefont {W.~T.}\ \bibnamefont {Eckenhoff}},
  \bibinfo {author} {\bibfnamefont {T.}~\bibnamefont {Pintauer}}, \ and\
  \bibinfo {author} {\bibfnamefont {R.~C.}\ \bibnamefont {Jin}},\ }\href@noop
  {} {\bibfield  {journal} {\bibinfo  {journal} {J. Phys. Chem. C},\ }\textbf
  {\bibinfo {volume} {112}},\ \bibinfo {pages} {14221} (\bibinfo {year}
  {2008}{\natexlab{b}})}\BibitemShut {NoStop}%
\bibitem [{\citenamefont {Chaki}\ \emph {et~al.}(2008)\citenamefont {Chaki},
  \citenamefont {Negishi}, \citenamefont {Tsunoyama}, \citenamefont
  {Shichibu},\ and\ \citenamefont {Tsukuda}}]{ChakiAu38}%
  \BibitemOpen
  \bibfield  {author} {\bibinfo {author} {\bibfnamefont {N.~K.}\ \bibnamefont
  {Chaki}}, \bibinfo {author} {\bibfnamefont {Y.}~\bibnamefont {Negishi}},
  \bibinfo {author} {\bibfnamefont {H.}~\bibnamefont {Tsunoyama}}, \bibinfo
  {author} {\bibfnamefont {Y.}~\bibnamefont {Shichibu}}, \ and\ \bibinfo
  {author} {\bibfnamefont {T.}~\bibnamefont {Tsukuda}},\ }\href@noop {}
  {\bibfield  {journal} {\bibinfo  {journal} {J. Am. Chem. Soc.},\ }\textbf
  {\bibinfo {volume} {130}},\ \bibinfo {pages} {8608} (\bibinfo {year}
  {2008})}\BibitemShut {NoStop}%
\bibitem [{\citenamefont {Qian}\ \emph
  {et~al.}(2010){\natexlab{a}}\citenamefont {Qian}, \citenamefont {Eckenhoff},
  \citenamefont {Zhu}, \citenamefont {Pintauer},\ and\ \citenamefont
  {Jin}}]{JinAu38}%
  \BibitemOpen
  \bibfield  {author} {\bibinfo {author} {\bibfnamefont {H.~F.}\ \bibnamefont
  {Qian}}, \bibinfo {author} {\bibfnamefont {W.~T.}\ \bibnamefont {Eckenhoff}},
  \bibinfo {author} {\bibfnamefont {Y.}~\bibnamefont {Zhu}}, \bibinfo {author}
  {\bibfnamefont {T.}~\bibnamefont {Pintauer}}, \ and\ \bibinfo {author}
  {\bibfnamefont {R.~C.}\ \bibnamefont {Jin}},\ }\href@noop {} {\bibfield
  {journal} {\bibinfo  {journal} {J. Am. Chem. Soc.},\ }\textbf {\bibinfo
  {volume} {132}},\ \bibinfo {pages} {8280} (\bibinfo {year}
  {2010}{\natexlab{a}})}\BibitemShut {NoStop}%
\bibitem [{\citenamefont {Qian}\ \emph
  {et~al.}(2010){\natexlab{b}}\citenamefont {Qian}, \citenamefont {Chu},\ and\
  \citenamefont {Jin}}]{Au40}%
  \BibitemOpen
  \bibfield  {author} {\bibinfo {author} {\bibfnamefont {H.~F.}\ \bibnamefont
  {Qian}}, \bibinfo {author} {\bibfnamefont {Y.}~\bibnamefont {Chu}}, \ and\
  \bibinfo {author} {\bibfnamefont {R.~C.}\ \bibnamefont {Jin}},\ }\href@noop
  {} {\bibfield  {journal} {\bibinfo  {journal} {J. Am. Chem. Soc.},\ }\textbf
  {\bibinfo {volume} {132}},\ \bibinfo {pages} {4583} (\bibinfo {year}
  {2010}{\natexlab{b}})}\BibitemShut {NoStop}%
\bibitem [{\citenamefont {Dass}(2009)}]{Au68}%
  \BibitemOpen
  \bibfield  {author} {\bibinfo {author} {\bibfnamefont {A.}~\bibnamefont
  {Dass}},\ }\href@noop {} {\bibfield  {journal} {\bibinfo  {journal} {J. Am.
  Chem. Soc.},\ }\textbf {\bibinfo {volume} {131}},\ \bibinfo {pages} {11666}
  (\bibinfo {year} {2009})}\BibitemShut {NoStop}%
\bibitem [{\citenamefont {Jadzinsky}\ \emph {et~al.}(2007)\citenamefont
  {Jadzinsky}, \citenamefont {Calero}, \citenamefont {Ackerson}, \citenamefont
  {Bushnell},\ and\ \citenamefont {Kornberg}}]{Science1}%
  \BibitemOpen
  \bibfield  {author} {\bibinfo {author} {\bibfnamefont {P.~D.}\ \bibnamefont
  {Jadzinsky}}, \bibinfo {author} {\bibfnamefont {G.}~\bibnamefont {Calero}},
  \bibinfo {author} {\bibfnamefont {C.~J.}\ \bibnamefont {Ackerson}}, \bibinfo
  {author} {\bibfnamefont {D.~A.}\ \bibnamefont {Bushnell}}, \ and\ \bibinfo
  {author} {\bibfnamefont {R.~D.}\ \bibnamefont {Kornberg}},\ }\href@noop {}
  {\bibfield  {journal} {\bibinfo  {journal} {Science},\ }\textbf {\bibinfo
  {volume} {318}},\ \bibinfo {pages} {430} (\bibinfo {year}
  {2007})}\BibitemShut {NoStop}%
\bibitem [{\citenamefont {Qian}\ and\ \citenamefont {Jin}(2009)}]{JinAu144}%
  \BibitemOpen
  \bibfield  {author} {\bibinfo {author} {\bibfnamefont {H.~F.}\ \bibnamefont
  {Qian}}\ and\ \bibinfo {author} {\bibfnamefont {R.~C.}\ \bibnamefont {Jin}},\
  }\href@noop {} {\bibfield  {journal} {\bibinfo  {journal} {Nano Lett.},\
  }\textbf {\bibinfo {volume} {9}},\ \bibinfo {pages} {4083} (\bibinfo {year}
  {2009})}\BibitemShut {NoStop}%
\bibitem [{\citenamefont {Fields-Zinna}\ \emph {et~al.}(2009)\citenamefont
  {Fields-Zinna}, \citenamefont {Sardar}, \citenamefont {Beasley},\ and\
  \citenamefont {Murray}}]{Murray}%
  \BibitemOpen
  \bibfield  {author} {\bibinfo {author} {\bibfnamefont {C.~A.}\ \bibnamefont
  {Fields-Zinna}}, \bibinfo {author} {\bibfnamefont {R.}~\bibnamefont
  {Sardar}}, \bibinfo {author} {\bibfnamefont {C.~A.}\ \bibnamefont {Beasley}},
  \ and\ \bibinfo {author} {\bibfnamefont {R.~W.}\ \bibnamefont {Murray}},\
  }\href@noop {} {\bibfield  {journal} {\bibinfo  {journal} {J. Am. Chem.
  Soc.},\ }\textbf {\bibinfo {volume} {131}},\ \bibinfo {pages} {16266}
  (\bibinfo {year} {2009})}\BibitemShut {NoStop}%
\bibitem [{\citenamefont {Akola}\ \emph {et~al.}(2008)\citenamefont {Akola},
  \citenamefont {Walter}, \citenamefont {Whetten}, \citenamefont
  {H{\"a}kkinen},\ and\ \citenamefont {Gr{\"o}nbeck}}]{AU25-01}%
  \BibitemOpen
  \bibfield  {author} {\bibinfo {author} {\bibfnamefont {J.}~\bibnamefont
  {Akola}}, \bibinfo {author} {\bibfnamefont {M.}~\bibnamefont {Walter}},
  \bibinfo {author} {\bibfnamefont {R.~L.}\ \bibnamefont {Whetten}}, \bibinfo
  {author} {\bibfnamefont {H.}~\bibnamefont {H{\"a}kkinen}}, \ and\ \bibinfo
  {author} {\bibfnamefont {H.}~\bibnamefont {Gr{\"o}nbeck}},\ }\href@noop {}
  {\bibfield  {journal} {\bibinfo  {journal} {J. Am. Chem. Soc},\ }\textbf
  {\bibinfo {volume} {130}},\ \bibinfo {pages} {3756} (\bibinfo {year}
  {2008})}\BibitemShut {NoStop}%
\bibitem [{\citenamefont {Akola}\ \emph {et~al.}(2010)\citenamefont {Akola},
  \citenamefont {Kacprzak}, \citenamefont {Lopez-Acevedo}, \citenamefont
  {Walter}, \citenamefont {Gr{\"o}nbeck},\ and\ \citenamefont
  {H{\"a}kkinen}}]{AU25-05}%
  \BibitemOpen
  \bibfield  {author} {\bibinfo {author} {\bibfnamefont {J.}~\bibnamefont
  {Akola}}, \bibinfo {author} {\bibfnamefont {K.~A.}\ \bibnamefont {Kacprzak}},
  \bibinfo {author} {\bibfnamefont {O.}~\bibnamefont {Lopez-Acevedo}}, \bibinfo
  {author} {\bibfnamefont {M.}~\bibnamefont {Walter}}, \bibinfo {author}
  {\bibfnamefont {H.}~\bibnamefont {Gr{\"o}nbeck}}, \ and\ \bibinfo {author}
  {\bibfnamefont {H.}~\bibnamefont {H{\"a}kkinen}},\ }\href@noop {} {\bibfield
  {journal} {\bibinfo  {journal} {J. Phys. Chem. C},\ }\textbf {\bibinfo
  {volume} {114}},\ \bibinfo {pages} {15986} (\bibinfo {year}
  {2010})}\BibitemShut {NoStop}%
\bibitem [{\citenamefont {Reveles}\ \emph {et~al.}(2009)\citenamefont
  {Reveles}, \citenamefont {Clayborne}, \citenamefont {Reber}, \citenamefont
  {Khanna}, \citenamefont {Pradhan}, \citenamefont {Sen},\ and\ \citenamefont
  {Pederson}}]{Khanna}%
  \BibitemOpen
  \bibfield  {author} {\bibinfo {author} {\bibfnamefont {J.~U.}\ \bibnamefont
  {Reveles}}, \bibinfo {author} {\bibfnamefont {P.~A.}\ \bibnamefont
  {Clayborne}}, \bibinfo {author} {\bibfnamefont {A.~C.}\ \bibnamefont
  {Reber}}, \bibinfo {author} {\bibfnamefont {S.~N.}\ \bibnamefont {Khanna}},
  \bibinfo {author} {\bibfnamefont {K.}~\bibnamefont {Pradhan}}, \bibinfo
  {author} {\bibfnamefont {P.}~\bibnamefont {Sen}}, \ and\ \bibinfo {author}
  {\bibfnamefont {M.~R.}\ \bibnamefont {Pederson}},\ }\href@noop {} {\bibfield
  {journal} {\bibinfo  {journal} {Nature Chemistry},\ }\textbf {\bibinfo
  {volume} {1}},\ \bibinfo {pages} {310} (\bibinfo {year} {2009})}\BibitemShut
  {NoStop}%
\bibitem [{\citenamefont {Mortensen}\ \emph {et~al.}(2005)\citenamefont
  {Mortensen}, \citenamefont {Hansen},\ and\ \citenamefont {Jacobsen}}]{GPAW1}%
  \BibitemOpen
  \bibfield  {author} {\bibinfo {author} {\bibfnamefont {J.~J.}\ \bibnamefont
  {Mortensen}}, \bibinfo {author} {\bibfnamefont {L.~B.}\ \bibnamefont
  {Hansen}}, \ and\ \bibinfo {author} {\bibfnamefont {K.~W.}\ \bibnamefont
  {Jacobsen}},\ }\href@noop {} {\bibfield  {journal} {\bibinfo  {journal}
  {Phys. Rev. B},\ }\textbf {\bibinfo {volume} {71}},\ \bibinfo {pages}
  {035109} (\bibinfo {year} {2005})}\BibitemShut {NoStop}%
\bibitem [{\citenamefont {Enkovaara}\ and\ \citenamefont
  {et~al}(2010)}]{GPAW2}%
  \BibitemOpen
  \bibfield  {author} {\bibinfo {author} {\bibfnamefont {J.}~\bibnamefont
  {Enkovaara}}\ and\ \bibinfo {author} {\bibnamefont {et~al}},\ }\href@noop {}
  {\bibfield  {journal} {\bibinfo  {journal} {J. Phys. Condens. Matter},\
  }\textbf {\bibinfo {volume} {22}},\ \bibinfo {pages} {253202} (\bibinfo
  {year} {2010})}\BibitemShut {NoStop}%
\bibitem [{\citenamefont {Perdew}\ \emph {et~al.}(1996)\citenamefont {Perdew},
  \citenamefont {Burke},\ and\ \citenamefont {Ernzerhof}}]{PBE}%
  \BibitemOpen
  \bibfield  {author} {\bibinfo {author} {\bibfnamefont {J.~P.}\ \bibnamefont
  {Perdew}}, \bibinfo {author} {\bibfnamefont {K.}~\bibnamefont {Burke}}, \
  and\ \bibinfo {author} {\bibfnamefont {M.}~\bibnamefont {Ernzerhof}},\
  }\href@noop {} {\bibfield  {journal} {\bibinfo  {journal} {Phys. Rev.
  Lett.},\ }\textbf {\bibinfo {volume} {77}},\ \bibinfo {pages} {3865}
  (\bibinfo {year} {1996})}\BibitemShut {NoStop}%
\bibitem [{\citenamefont {Tang}\ \emph {et~al.}(2009)\citenamefont {Tang},
  \citenamefont {Sanville},\ and\ \citenamefont {Henkelman}}]{Bader2}%
  \BibitemOpen
  \bibfield  {author} {\bibinfo {author} {\bibfnamefont {W.}~\bibnamefont
  {Tang}}, \bibinfo {author} {\bibfnamefont {E.}~\bibnamefont {Sanville}}, \
  and\ \bibinfo {author} {\bibfnamefont {G.}~\bibnamefont {Henkelman}},\
  }\href@noop {} {\bibfield  {journal} {\bibinfo  {journal} {J. Phys. Condens.
  Matter},\ }\textbf {\bibinfo {volume} {21}},\ \bibinfo {pages} {084204}
  (\bibinfo {year} {2009})}\BibitemShut {NoStop}%
\bibitem [{\citenamefont {Walter}\ \emph {et~al.}(2008)\citenamefont {Walter},
  \citenamefont {Akola}, \citenamefont {Lopez-Acevedo}, \citenamefont
  {Jadzinsky}, \citenamefont {Calero}, \citenamefont {Ackerson}, \citenamefont
  {Whetten}, \citenamefont {Gr{\"o}nbeck},\ and\ \citenamefont
  {H{\"a}kkinen}}]{PNAS1}%
  \BibitemOpen
  \bibfield  {author} {\bibinfo {author} {\bibfnamefont {M.}~\bibnamefont
  {Walter}}, \bibinfo {author} {\bibfnamefont {J.}~\bibnamefont {Akola}},
  \bibinfo {author} {\bibfnamefont {O.}~\bibnamefont {Lopez-Acevedo}}, \bibinfo
  {author} {\bibfnamefont {P.~D.}\ \bibnamefont {Jadzinsky}}, \bibinfo {author}
  {\bibfnamefont {G.}~\bibnamefont {Calero}}, \bibinfo {author} {\bibfnamefont
  {C.~J.}\ \bibnamefont {Ackerson}}, \bibinfo {author} {\bibfnamefont {R.~L.}\
  \bibnamefont {Whetten}}, \bibinfo {author} {\bibfnamefont {H.}~\bibnamefont
  {Gr{\"o}nbeck}}, \ and\ \bibinfo {author} {\bibfnamefont {H.}~\bibnamefont
  {H{\"a}kkinen}},\ }\href@noop {} {\bibfield  {journal} {\bibinfo  {journal}
  {Proc. Natl. Acad. Sci (USA)},\ }\textbf {\bibinfo {volume} {105}},\ \bibinfo
  {pages} {9157} (\bibinfo {year} {2008})}\BibitemShut {NoStop}%
\bibitem [{\citenamefont {Zhou}\ \emph {et~al.}(2011)\citenamefont {Zhou},
  \citenamefont {Cai}, \citenamefont {Zeng}, \citenamefont {Zhang},\ and\
  \citenamefont {Feng}}]{Feng}%
  \BibitemOpen
  \bibfield  {author} {\bibinfo {author} {\bibfnamefont {M.}~\bibnamefont
  {Zhou}}, \bibinfo {author} {\bibfnamefont {Y.~Q.}\ \bibnamefont {Cai}},
  \bibinfo {author} {\bibfnamefont {M.~G.}\ \bibnamefont {Zeng}}, \bibinfo
  {author} {\bibfnamefont {C.}~\bibnamefont {Zhang}}, \ and\ \bibinfo {author}
  {\bibfnamefont {Y.~P.}\ \bibnamefont {Feng}},\ }\href@noop {} {\bibfield
  {journal} {\bibinfo  {journal} {Appl. Phys. Lett},\ }\textbf {\bibinfo
  {volume} {98}},\ \bibinfo {pages} {143103} (\bibinfo {year}
  {2011})}\BibitemShut {NoStop}%
\end{thebibliography}%

\end{document}